\documentclass{pasj02}
\usepackage{graphicx}
\usepackage{bm}
\usepackage{natbib}
\usepackage{mathrsfs}
\usepackage{comment}
\usepackage{color}

\newcommand{\mr}[1]{\mathrm{#1}}
\newcommand{\msb}[1]{_{\mathrm{#1}}}

\newcommand{\brp}[1]{\left(#1\right)}
\newcommand{\brb}[1]{\left[#1\right]}

\def\arcs{\,\mathrm{arcsec}}

\def\qn{q_{\nu}}
\def\sfr{{\rm SFR}}
\def\msun{{\rm M}_\odot}

\usepackage[switch,mathlines]{lineno}
\begin{document}

\Received{}%{yyyy/mm/dd}
\Accepted{}%{yyyy/mm/dd}
%\Published{yyyy/mm/dd}

\title{{Estimating Galaxy Star Formation Rates from Murchison Widefield Array Radio Continuum Emission}}

%% begin:list of authors
% Do NOT capitalize all letters in "textsc".
\author{%
Tsutomu T. \textsc{Takeuchi}\altaffilmark{1,2}	
Suchetha \textsc{Cooray}\altaffilmark{3,4,5},
Luca \textsc{Cortese}\altaffilmark{6,7},
O.\ Ivy \textsc{Wong}\altaffilmark{8}, 
Barbara \textsc{Catinella}\altaffilmark{6,7}, 
and
Shuntaro A.\ \textsc{Yoshida}\altaffilmark{1}
}

\altaffiltext{1}{Division of Particle and Astrophysical Science, Nagoya University, Furo-cho, Chikusa-ku, Nagoya, 464--8602, Japan}
\altaffiltext{2}{The Research Center for Statistical Machine
Learning, the Institute of Statistical Mathematics, 10--3 Midori-cho, Tachikawa, Tokyo 190--8562, Japan}
\altaffiltext{{3}}{{Kavli Institute for Particle Astrophysics and Cosmology, Stanford University, 452 Lomita Mall, Stanford, CA 94305, USA}}
\altaffiltext{{4}}{{SLAC National Accelerator Laboratory, 2575 Sand Hill Road, Menlo Park, CA 94025, USA}}
\altaffiltext{{5}}{{Astronomical Institute, Tohoku University, Aramaki, Aoba-ku, Sendai, Miyagi 980-8578, Japan}}
\altaffiltext{6}{International Centre for Radio Astronomy Research (ICRAR), University of Western Australia, Crawley, Perth, Western Australia 6009, Australia}
\altaffiltext{7}{ARC Centre of Excellence for All Sky Astrophysics in 3 Dimensions (ASTRO 3D), Australia}
\altaffiltext{8}{CSIRO Space \& Astronomy, PO Box 1130, Bentley, Western Australia 6102, Australia}

\email{tsutomu.takeuchi.ttt@gmail.com}


\maketitle
\begin{abstract}
{
We investigate the frequency dependence of the low-frequency infrared (IR)--radio correlation for nearby galaxies drawn from the {\sl Herschel} Reference Survey (HRS), using radio continuum observations from the GaLactic and Extragalactic All-sky MWA (GLEAM) survey obtained with the Murchison Widefield Array (MWA). 
The GLEAM survey provides densely sampled flux densities across 20 sub-bands spanning {$72\mbox{--}231$~MHz}, allowing detailed characterization of the low-frequency radio spectral energy distribution (SED). 
We analyze 18 nearby galaxies detected by GLEAM and examine the frequency dependence of the IR--radio correlation coefficient $q_\nu$. 
Among these galaxies, three sources show Seyfert or LINER signatures and may contain non-negligible AGN contributions to the radio continuum emission. 
We therefore use the remaining 15 predominantly star-forming galaxies as the primary sample for deriving the main calibration results. 
For the majority of the sample, the low-frequency radio continuum is well described by a single power-law spectral model from the GLEAM frequency range up to {$1.5$~GHz}, with a median spectral index of $\gamma=-0.63$. 
The ensemble-averaged $q_\nu$ relation and its intrinsic scatter provide a practical calibration for converting low-frequency radio luminosities into star formation rates (SFRs), while explicitly quantifying the associated systematic uncertainty. 
Our results indicate that the low-frequency IR--radio correlation remains stable across the MWA bands for nearby star-forming galaxies, supporting the use of low-frequency radio continuum as a dust-unbiased tracer of star formation in the local Universe. 
These results provide a local benchmark for interpreting low-frequency radio continuum surveys and for future studies of radio-based SFR tracers with next-generation facilities such as the Square Kilometre Array (SKA).}
\end{abstract}

\section{Introduction}\label{sec:introduction}

Star formation is one of the most fundamental aspects of the evolution of galaxies. 
Observationally, the only viable path to estimate the star formation rate (SFR) in a galaxy is to detect its electromagnetic flux (or flux density), convert this to luminosity, and relate the luminosity to the number of ionizing photons produced by massive stars.
We can then estimate the number of OB stars through stellar spectral energy distribution (SED) models \citep[e.g.,][]{1983ApJ...272...54K}.
This method is based on the short lifetime of OB stars ($\sim 10^6$ yr) compared with the typical age of galaxies ($\sim 10^{10}$ yr), allowing us to use OB stars as ``instantaneous'' tracers of recent star formation under an assumed initial mass function \citep[e.g.,][]{2002Sci...295...82K}.

Thus, the crucial point is to measure observables directly connected to OB stars.
Because of their high luminosity and temperature, ionizing ultraviolet (UV) radiation is the most straightforward indicator. 
Hydrogen recombination lines such as H$\alpha$ provide strong diagnostics \citep[e.g.,][]{Kennicutt1998}.
Non-ionizing UV continuum is also widely used \citep[e.g.,][]{2005A&A...440L..17T,2007ApJS..173..404B,2010A&A...514A...4T}, though its sensitivity to dust obscuration limits its reliability when used alone.
Consequently, mid- and far-infrared (MIR and FIR) emission, which traces dust heated by OB stars, has become a powerful complementary estimator of SFR \citep[e.g.,][]{Kennicutt1998,2005A&A...440L..17T,2010A&A...514A...4T,2013EP&S...65..281T}.
To mitigate biases intrinsic to individual tracers, hybrid UV+IR indicators are now widely adopted \citep[e.g.,][]{Kennicutt2009,2010A&A...514A...4T,2011ApJ...737...67M}.  
Other indirect SFR indicators (e.g., X-rays, nebular forbidden lines, PAH emission) have also been proposed \citep[e.g.,][]{Kennicutt1998}.

Among these methods, long-wavelength radio continuum has emerged as one of the most promising probes, particularly for high-redshift galaxy evolution \citep{2000ApJ...544..641H,2011ApJ...737...67M,2016arXiv160301938T}.
Radio continuum emission is primarily composed of free--free emission from H{\sc ii} regions and synchrotron emission from relativistic electrons spiraling in interstellar magnetic fields \citep[][]{1992ARA&A..30..575C}.
{Free--free emission is directly linked to ionizing massive stars, whereas synchrotron emission traces cosmic-ray electrons accelerated by supernova remnants and is therefore affected by transport processes and time delays associated with cosmic-ray propagation \citep[e.g.,][]{2024MNRAS.531..708C}. 
Nevertheless, both components are ultimately connected to recent star formation activity and contribute to the observed IR--radio correlation.}
A major advantage of radio continuum is its immunity to dust extinction, making it a powerful probe of obscured star formation activity.
With the advent of the Square Kilometre Array (SKA), radio continuum observations will become increasingly important for tracing galaxy evolution and the cosmic star formation history \citep[e.g.,][]{2015aska.confE..85M,2016arXiv160301938T}. 

However, the calibration of radio continuum as a quantitative SFR tracer remains essentially empirical, unlike H$\alpha$ or UV calibrations \citep[][]{2011ApJ...737...67M}. 
This calibration is typically anchored in the IR--radio correlation \citep[e.g.,][]{1992ARA&A..30..575C}, which has been extensively investigated using {\sl IRAS}, {\sl ISO}, {\sl Spitzer}, {\sl AKARI}, and {\sl Herschel} observations \citep[e.g.,][]{1986A&A...160L...4H,1989ApJ...340..708D,1990ApJ...362...59B,2003A&A...409..907P,2004ApJS..154..147A,2006ApJ...638..157M,2010MNRAS.409...92J,2014MNRAS.445.2232S,2019PASJ...71...28S}. 
Although many studies have attempted to explore the redshift evolution of the IR--radio correlation, the physical origin of its scatter and frequency dependence remains incompletely understood \citep[e.g.,][]{2011MNRAS.410.1155B,2011ApJ...731...79M,2013A&A...556A.142S,2016ApJ...827..109S,2021A&A...647A.123D,2021MNRAS.504..118M}. 

In the past several years, LOFAR observations have substantially advanced our understanding of low-frequency radio emission in star-forming galaxies. 
Large samples from the LOFAR Two-metre Sky Survey (LoTSS) enabled detailed investigations of the frequency dependence and scatter of the low-frequency IR--radio correlation across broad redshift ranges \citep[e.g.,][]{2018MNRAS.475.3010G,2019A&A...631A.109W,2021A&A...648A...6S}. 
Broadband radio analyses further revealed the importance of spectral curvature, including low-frequency flattening and turnover caused by free--free absorption and cosmic-ray electron energy losses \citep[e.g.,][]{2024MNRAS.528.5346A,2025PASA...42....2G}. 
These studies highlight the need for accurate characterization of the radio spectral index $\gamma$ and the frequency dependence of the IR--radio correlation at low radio frequencies.

The Murchison Widefield Array (MWA) provides a unique opportunity for such studies through the GLEAM survey \citep{Wayth2015,Hurley-Walker2017a}, which delivers densely sampled radio continuum measurements across 20 narrow sub-bands between 72 and 231\,MHz. 
Unlike single-frequency surveys, this dense frequency coverage enables detailed characterization of the low-frequency radio spectral energy distribution (SED) and the frequency dependence of the IR--radio correlation. 

In this work, we analyze 18 nearby galaxies from the {\sl Herschel} Reference Survey (HRS) detected by GLEAM, with the aim of:
(1) characterizing the frequency dependence of the IR--radio correlation;
(2) measuring the low-frequency radio spectral index across the GLEAM bands; and
{(3) examining the implications of the observed $q_\nu$ relation for radio-based SFR calibration.}
{Because a few galaxies exhibit Seyfert or LINER signatures in optical diagnostics, we additionally examine the robustness of the derived relations using a reduced subsample excluding these objects.}
Our study complements LOFAR-based statistical investigations by providing a nearby benchmark sample with well-characterized multi-wavelength data. 
{In particular, the detailed source properties summarized in Table~\ref{tab:sample}, including morphology, optical size, distance, and AGN classification, allow us to assess possible effects of source blending, angular resolution, and nuclear activity on the low-frequency IR--radio correlation.}
This enables a detailed examination of the physical and observational properties underlying low-frequency radio continuum measurements in nearby galaxies.

{T}his paper is organized as follows. 
We introduce the properties of the sample galaxies in Section~\ref{sec:sample}. 
{In Section~\ref{sec:method}, we describe the methodology used to characterize the frequency dependence of the IR--radio correlation and its implications for radio-based SFR calibration.}
The results are presented in Section~\ref{sec:results}, and discussed further in Section~\ref{sec:discussion}. 
Section~\ref{sec:summary} summarizes the paper. 
Images and radio SED fitting results are presented in the Appendix~\ref{sec:galaxy_image} and \ref{sec:fitting_result}.

\section{The Sample}\label{sec:sample}

\begin{table*}
	\centering
    \caption{
        The 18 HRS galaxies selected in Section~\ref{sec:sample}.
        Column (1): HRS ID; (2): NGC number; (3): R.A.; (4): Dec.; (5): morphological type from \citet{Ciesla2014} 
        (except for HRS 163, taken from \citealt{Cortese2012}); 
        (6): $D_{25}$ and (7): distances from \citet{Cortese2012}; 
        (8): GLEAM ID; and (9): AGN classification.
        The ``AGN'' column is based on optical diagnostics using the BPT diagram 
        \citep[e.g.][]{Baldwin1981, Kewley2001, Kauffmann2003, Schawinski2007} 
        and emission-line measurements from \citet{Boselli2015}.
        {The three galaxies classified as Seyfert or LINER systems (HRS~144, HRS~163, and HRS~220) are retained in this table for completeness, but are excluded from the clean star-forming sample used for the primary calibration analysis.}
    }\label{tab:sample}
\begin{tabular}{lllllrrlll}
\hline
\hline
HRS ID &   NGC &         R.A. &          Dec &      Type &   D25 &  Dist. &        GLEAM ID & AGN \\%& HI-def \\
 &      &              &              &           &  [arcmin] &  [Mpc] &             &              \\%           &              \\
\hline
HRS	25	&	3437	&	  10:52:35.75 	&	  $+$22:56:02.9 	&	        Sc 	&	2.51	&	18.24	&	  J105236$+$225606 	&	     -- 	\\
HRS	36	&	3504	&	  11:03:11.21 	&	  $+$27:58:21.0 	&	       Sab 	&	2.69	&	21.94	&	  J110311$+$275812 	&	     -- 	\\
HRS	50	&	3655	&	  11:22:54.62 	&	  $+$16:35:24.5 	&	        Sc 	&	1.55	&	21.43	&	  J112254$+$163522 	&	     -- 	\\
HRS	77	&	4030	&	  12:00:23.64 	&	  $-$01:06:00.0 	&	       Sbc 	&	4.17	&	20.83	&	  J120023$-$010607 	&	     -- 	\\
HRS	102	&	4254	&	  12:18:49.63 	&	  $+$14:24:59.4 	&	        Sc 	&	6.15	&	17	&	  J121850$+$142515 	&	     -- 	\\
HRS	114	&	4303	&	  12:21:54.90 	&	  $+$04:28:25.1 	&	       Sbc 	&	6.59	&	17	&	  J122154$+$042827 	&	     -- 	\\
HRS	122	&	4321	&	  12:22:54.90 	&	  $+$15:49:20.6 	&	       Sbc 	&	9.12	&	17	&	  J122255$+$154939 	&	     -- 	\\
HRS	144	&	4388	&	  12:25:46.82 	&	  $+$12:39:43.5 	&	        Sb 	&	5.1	&	17	&	  J122548$+$123917 	&	  Seyfert 	\\
HRS	163	&	4438	&	  12:27:45.59 	&	  $+$13:00:31.8 	&	        Sb 	&	8.12	&	17	&	  J122744$+$130020 	&	  Seyfert 	\\
HRS	190	&	4501	&	  12:31:59.22 	&	  $+$14:25:13.5 	&	        Sb 	&	7.23	&	17	&	  J123159$+$142503 	&	     -- 	\\
HRS	201	&	4527	&	  12:34:08.50 	&	  $+$02:39:13.7 	&	       Sbc 	&	5.86	&	17	&	  J123408$+$023909 	&	     -- 	\\
HRS	203	&	4532	&	  12:34:19.33 	&	  $+$06:28:03.7 	&	  Im (Im/S) 	&	2.6	&	17	&	  J123420$+$062758 	&	     -- 	\\
HRS	204	&	4535	&	  12:34:20.31 	&	  $+$08:11:51.9 	&	        Sc 	&	8.33	&	17	&	  J123418$+$081157 	&	     -- 	\\
HRS	205	&	4536	&	  12:34:27.13 	&	  $+$02:11:16.4 	&	       Sbc 	&	7.23	&	17	&	  J123427$+$021114 	&	     -- 	\\
HRS	220	&	4579	&	  12:37:43.52 	&	  $+$11:49:05.5 	&	        Sb 	&	6.29	&	17	&	  J123743$+$114909 	&	  LINER 	\\
HRS	247	&	4654	&	  12:43:56.58 	&	  $+$13:07:36.0 	&	       Scd 	&	4.99	&	17	&	  J124355$+$130801 	&	     -- 	\\
HRS	251	&	4666	&	  12:45:08.59 	&	  $-$00:27:42.8 	&	        Sc 	&	4.57	&	21.61	&	  J124508$-$002747 	&	     -- 	\\
HRS	306	&	5363	&	  13:56:07.21 	&	  $+$05:15:17.2 	&	       pec 	&	4.07	&	16.23	&	  J135607$+$051516 	&	     -- 	\\
\hline
\end{tabular}
\end{table*}

The {\sl Herschel} Reference Survey (HRS) \citep{2010PASP..122..261B} contains 322 nearby galaxies observed across a wide wavelength range, from the far-ultraviolet (FUV) to the radio at $1500\;\mathrm{MHz}$ \citep[e.g.,][]{Cortese2012, Ciesla2014, Boselli2015}.
The survey is volume-limited, including galaxies with distances between 15 and 25~Mpc.
All late-type spirals and irregulars (Sa--Sd--Im--BCD) with 2MASS $K$-band total magnitude $K_{\rm Stot} \leq 12$, as well as ellipticals and lenticulars (E, S0, S0a) with $K_{\rm Stot} \leq 8.7$, were selected.
This $K$-band criterion minimizes biases associated with young stellar populations and dust attenuation, providing an approximately stellar-mass-selected sample.
The HRS galaxies span a broad range of morphologies (early-type, late-type, and peculiar) and environments, from the Virgo cluster core to isolated field regions.
Thus, they form an excellent reference sample for studying representative properties of local galaxies.

The GaLactic Extragalactic All-sky MWA (GLEAM) survey \citep{Wayth2015, Hurley-Walker2017a} covers most of the southern sky and the northern sky up to $+30^{\circ}$ (approximately $25,000\,\mathrm{deg}^2$), using the Murchison Widefield Array \citep[MWA;][]{Tingay2013a} in Western Australia.
GLEAM provides flux densities in 20 narrow sub-bands spanning $72\mbox{--}231\;\mathrm{MHz}$ (each with $7.6\;\mathrm{MHz}$ bandwidth), with typical sensitivity of $\sim 7\;\mathrm{mJy}$ and angular resolution of $\sim 2\,\mathrm{arcmin}$ at $200\;\mathrm{MHz}$.
This dense low-frequency sampling is particularly advantageous for reconstructing radio SEDs and measuring spectral indices.
{For comparison, the $1500\;\mathrm{MHz}$ radio flux densities used below are taken from the compilation of \citet{Boselli2015}; the relevant VLA/NVSS angular resolution is approximately $45''$.}
{The difference between the GLEAM and $1500\;\mathrm{MHz}$ angular resolutions is considered when interpreting the fitted radio spectral indices, because resolution mismatch can affect integrated flux measurements for nearby extended galaxies.}

After cross-matching the HRS and GLEAM catalogs within a radius of $120''$, we identify 39 HRS galaxies with a potential radio counterpart in GLEAM.
{The matching radius was chosen to be comparable to the GLEAM beam size at $200\;\mathrm{MHz}$.}
{Because HRS galaxies are nearby and often spatially extended, a substantially smaller radius could miss genuine low-frequency counterparts associated with diffuse radio emission.}
{At the same time, this relatively large radius increases the possibility of source confusion; therefore, we visually inspected the radio contours overlaid on the optical images and removed blended sources in the first step below.}
We select our final subsample through the following steps:
\begin{enumerate}
    \item Remove blended sources ($39 \rightarrow 33$);
    \item Remove elliptical galaxies ($33 \rightarrow 29$);
    \item Select objects with at least two high-quality MWA flux measurements ($\mathrm{SNR} > 5$), yielding $29 \rightarrow 18$.
\end{enumerate}
{Step (1) removes sources whose radio emission cannot be reliably deblended at MWA resolution.}
{Step (2) excludes elliptical galaxies, whose radio emission is generally dominated by AGN-related processes rather than star formation.}
{Step (3) ensures that the radio SED can be meaningfully constrained within the MWA band.}

This procedure yielded a full GLEAM-detected HRS sample of 18 nearby galaxies.
{Among these 18 galaxies, three objects are classified as Seyfert or LINER systems based on optical emission-line diagnostics: HRS~144, HRS~163, and HRS~220.}
{Because even weak AGN activity can contribute to the radio continuum through compact cores or low-power jets, we exclude these three objects from the primary calibration analysis.}
{We therefore define the remaining 15 galaxies as the clean star-forming sample and use this clean sample for the main derivation of the frequency-dependent IR--radio relation and radio-based SFR calibration.}
{The full 18-object sample is retained in Table~\ref{tab:sample} and used for transparency, for identifying possible AGN-affected systems, and for comparison with previous versions of the analysis.}
Their basic properties are summarized in Table~\ref{tab:sample}, and their radio contours overlaid on SDSS $i$-band images are shown in Figure~\ref{fig:galaxy_image} in Appendix~\ref{sec:galaxy_image}.

\section{Method: IR--Radio Correlation}\label{sec:method}

{The primary goal of this work is to characterize the frequency dependence and scatter of the low-frequency IR--radio correlation in nearby galaxies.}
To connect the low-frequency radio continuum emission to galaxy star formation activity, we use the IR emission as a reference tracer of recent star formation.
A tight correlation between the $60\,\mu{\rm m}$ and $1.4\;\mbox{GHz}$ flux densities was originally reported by \citet{Helou1985} and \citet{Condon1991a}.
They introduced the parameter $q$ as the logarithmic IR-to-radio ratio.
{In the present work, we generalize this quantity to multiple radio frequencies through $q_\nu$, and use it to investigate the frequency dependence of the low-frequency IR--radio correlation across the GLEAM bands.}

\subsection{IR-to-radio luminosity ratio $\qn$}\label{subsec:qvalue}

{We first define the frequency-dependent IR--radio correlation coefficient used throughout this paper.}
In this work, we adopt the following definition of $\qn$ \citep{CalistroRivera2017a}:
\begin{align}\label{eq:q_def}
    \qn &\equiv \log\brp{\frac{L_{8-1000\,\mu{\rm m}}}{3.75 \times 10^{12}\;\mr{erg\,s^{-1}\,Hz^{-1}}}} 
          - \log\brp{\frac{L_{\mr{radio},\nu}}{\mr{erg\,s^{-1}\,Hz^{-1}}}} ,
\end{align}
where $L_{8-1000\,\mu{\rm m}}$ is the total rest-frame IR luminosity integrated over $8\mbox{--}1000\,\mu{\rm m}$, which traces the total dust emission \citep[e.g.,][]{2005A&A...432..423T}, and $3.75 \times 10^{12}$ corresponds to the frequency of $80\,\mu{\rm m}$ and is introduced to make the quantity dimensionless.
$L_{\mr{radio},\nu}$ is the radio luminosity at frequency $\nu$.
{Throughout this paper, we use luminosities rather than flux densities in the definition of $q_\nu$.}

{Previous studies reported typical values of the IR--radio correlation coefficient of}
$q_{1400\,{\rm MHz}} = 2.14 \pm 0.14$ {for local star-forming galaxies} \citep{Helou1985}
{and}
$q_{1400\,{\rm MHz}} = 2.34 \pm 0.26$ \citep{Yun2001a}.
\citet{2003ApJ...586..794B} showed that $q_{1400\,{\rm MHz}} = 2.64 \pm 0.26$ when the total IR emission $\brp{8\mbox{--}1000\,\mu{\rm m}}$ is used instead of the FIR emission.
At lower frequencies, \citet{CalistroRivera2017a} and \citet{Wang2019} found
$q_{151\,{\rm MHz}} = 1.72 \pm 0.529$ and $q_{151\,{\rm MHz}} = 2.14 \pm 0.34$, respectively.
{These studies suggest that the low-frequency IR--radio correlation remains broadly consistent with that observed at GHz frequencies, although with a larger intrinsic scatter.}

We derive $L_{8-1000\,\mu{\rm m}}$ using the calibration presented by \citet{Galametz2013} with IR fluxes from {\sl Spitzer} and {\sl Herschel}: PACS ($24,\,70,\,100,\,160\,\mu{\rm m}$; \citealt{Bendo2012, Cortese2014c}) and SPIRE ($250\,\mu{\rm m}$; \citealt{Ciesla2012}).
Although \citet{Ciesla2014} already estimated the total IR luminosities for most HRS galaxies using SED fitting, we recalculate $L_{8-1000\,\mu{\rm m}}$ from the discrete broadband IR flux densities \citep{2005A&A...432..423T, Galametz2013} because one of the flux densities is missing for HRS~163 in our sample.
For consistency, we adopt the recalculated $L_{8-1000\,\mu{\rm m}}$ values for all galaxies in the sample.
{We confirmed that our recalculated IR luminosities are consistent with the SED-based estimates of \citet{Ciesla2014} within the quoted uncertainties and do not exhibit any systematic offset.}
The impact of the method used to derive the total IR luminosity can be assessed from Figure~2 of \citet{Ciesla2014}.
Our IR luminosity calibration therefore does not affect the conclusions of this work.

\subsection{Frequency dependence of the IR--radio correlation}\label{subsec:SFR_radio}

{We now examine the frequency dependence of the IR--radio correlation across the GLEAM bands.}
We assume a simple power-law form and fit the following equation to the data at the MWA frequencies and at $1500\;\mbox{MHz}$ \citep{Boselli2015}:
\begin{align}
    \qn = \gamma\log{\nu} + \beta ,\label{eq:q_fitting}
\end{align}
where $\gamma$ is the power-law index and $\beta$ is the intercept.
{This parameterization characterizes the frequency dependence of the IR--radio correlation coefficient $q_\nu$, rather than the radio SED itself.}
In Section~\ref{subsec:GammaDistribution}, we present the distribution of the estimated $\gamma$ values for our sample.

{To connect the observed $q_\nu$ relation with radio-based SFR estimates, we further introduce the following relations:}
\begin{align}
    \mr{SFR}\msb{IR} &= 3.88 \times 10^{-44} L_{8-1000\,\mu{\rm m}}\;\mr{erg}\,{\rm s}^{-1} \; , \label{eq:sfrir}\\
    q_{\nu}
    &=
    q_{1500\,{\rm MHz}}
    +
    \gamma
    \log
    \brp{
        \frac{\nu}
             {1500\;{\rm MHz}}
    } . \label{eq:q_nuto1500}
\end{align}
Equation~(\ref{eq:sfrir}) gives the SFR estimated from the total IR emission \citep{2011ApJ...737...67M}, and equation~(\ref{eq:q_nuto1500}) relates $\qn$ at a given frequency $\nu$ to its value at $1500\;\mbox{MHz}$.

Substituting equations~(\ref{eq:q_def}) and~(\ref{eq:q_nuto1500}) into equation~(\ref{eq:sfrir}) yields the following expression linking the SFR to the radio luminosity at frequency $\nu$:
\begin{align}\label{eq:sfrfromradio}
    &\mr{SFR}_{\mr{radio},\,\nu} \nonumber \\
    &\qquad = \left( 1.46\times10^{-31}\right) 
      10^{q_{1500\,{\rm MHz}}} 
      {\brp{\frac{\nu}{1500\;\mbox{MHz}}}}^{\gamma} 
      L_{\mr{radio},\,\nu} .
\end{align}
{In the following sections, we primarily focus on the ensemble-averaged behavior and scatter of the low-frequency IR--radio correlation.
Figure~\ref{fig:SFR_ratio_comparison} illustrates how the derived $q_\nu$ relation can be used to connect low-frequency radio luminosities with commonly used SFR estimators.}

\section{Results}\label{sec:results}

\begin{table*}
%	\centering
    \caption{
        Estimated properties of the 18 HRS galaxies in our sample. 
        Columns are: (1) $\gamma\msb{MWA}$; (2) $\gamma\msb{MWA+1500\,MHz}$; 
        (3) $\mr{SFR\msb{radio,\,151\;\mbox{MHz}}}$; (4) $\mr{SFR}\msb{IR}$; and (5) $\sfr_{{\rm FUV + 24}\mu{\rm m}}$.}\label{tab:estimated_properties}
    \begin{center}
\begin{tabular}{lrcccc}
\hline
\hline
HRS ID &  \multicolumn{1}{c}{$\gamma_{\mathrm{MWA}}$} & $\gamma_{\mathrm{MWA+1500MHz}}$ & $\sfr_{\mathrm{radio}}$ & $\sfr_{\mathrm{IR}}$ & $\sfr_{{\rm FUV + 24}\mu{\rm m}}$  \\
       &                          &                                 & $\brb{\msun\,\mr{yr}^{-1}}$ & $\brb{\msun\,\mr{yr}^{-1}}$ & $\brb{\msun\,\mr{yr}^{-1}}$ \\
\hline
HRS	25	& $	 0.33 \pm 0.47 	$ & $	-0.66 \pm 0.07 	$ & $	0.96 \pm 0.28 	$ & $	  1.41 \pm 0.08 	$ & $	1.50 	$ \\
HRS	36	& $	-0.42 \pm 0.10 	$ & $	-0.53 \pm 0.05 	$ & $	5.15 \pm 1.35 	$ & $	  4.09 \pm 0.16 	$ & $	6.25 	$ \\
HRS	50	& $	-0.38 \pm 0.20 	$ & $	-0.63 \pm 0.04 	$ & $	1.45 \pm 0.41 	$ & $	  1.91 \pm 0.07 	$ & $	1.59 	$ \\
HRS	77	& $	-0.57 \pm 0.07 	$ & $	-0.63 \pm 0.04 	$ & $	3.55 \pm 0.94 	$ & $	  4.81 \pm 0.19 	$ & $	3.96 	$ \\
HRS	102	& $	-0.70 \pm 0.05 	$ & $	-0.70 \pm 0.03 	$ & $	8.19 \pm 2.12 	$ & $	  6.47 \pm 0.25 	$ & $	6.37 	$ \\
HRS	114	& $	-0.57 \pm 0.05 	$ & $	-0.58 \pm 0.04 	$ & $	5.49 \pm 1.43 	$ & $	  6.15 \pm 0.23 	$ & $	6.63 	$ \\
HRS	122	& $	-0.77 \pm 0.07 	$ & $	\mbox{--}       $ & $	5.45 \pm 0.22 	$ & $	  4.28 \pm 1.14 	$ & $	5.35 	$ \\
HRS	144	& $	 0.22 \pm 0.22 	$ & $	-0.73 \pm 0.08 	$ & $	3.09 \pm 0.85 	$ & $	  1.66 \pm 0.06 	$ & $	2.95 	$ \\
HRS	163	& $	-0.81 \pm 0.35 	$ & $	\mbox{--}      	$ & $	0.69 \pm 0.03 	$ & $     2.13 \pm 0.66 	$ & $	0.39 	$ \\
HRS	190	& $	-0.67 \pm 0.12 	$ & $	-0.72 \pm 0.05 	$ & $	4.92 \pm 1.31 	$ & $	  4.37 \pm 0.18 	$ & $	3.16 	$ \\
HRS	201	& $	-0.54 \pm 0.06 	$ & $	-0.56 \pm 0.04 	$ & $	2.71 \pm 0.71 	$ & $	  4.55 \pm 0.24 	$ & $	3.79 	$ \\
HRS	203	& $	-0.55 \pm 0.13 	$ & $	-0.60 \pm 0.04 	$ & $	1.52 \pm 0.43 	$ & $	  0.99 \pm 0.04 	$ & $	1.33 	$ \\
HRS	204	& $	-0.71 \pm 0.10 	$ & $	\mbox{--}      	$ & $	2.60 \pm 0.11 	$ & $     2.69 \pm 0.75 	$ & $	2.85 	$ \\
HRS	205	& $	-0.61 \pm 0.04 	$ & $	-0.57 \pm 0.02 	$ & $	2.63 \pm 0.69 	$ & $	  3.58 \pm 0.14 	$ & $	4.51 	$ \\
HRS	220	& $	-0.42 \pm 0.18 	$ & $	-0.77 \pm 0.05 	$ & $	2.32 \pm 0.65 	$ & $	  1.56 \pm 0.08 	$ & $	1.11 	$ \\
HRS	247	& $	-0.27 \pm 0.29 	$ & $	-0.66 \pm 0.06 	$ & $	1.65 \pm 0.49 	$ & $	  2.59 \pm 0.11 	$ & $	2.52 	$ \\
HRS	251	& $	-0.58 \pm 0.03 	$ & $	-0.58 \pm 0.02 	$ & $	9.11 \pm 2.35 	$ & $	  9.20 \pm 0.33 	$ & $	6.66 	$ \\
HRS	306	& $	-0.61 \pm 0.08 	$ & $	-0.51 \pm 0.04 	$ & $	1.75 \pm 0.46 	$ & $	  0.27 \pm 0.03 	$ & $	0.18 	$ \\
\hline
\end{tabular}
\end{center}
\end{table*}

We now present the results of our analysis.
All estimated quantities used in this work are summarized in Table~\ref{tab:estimated_properties}. 

\subsection{Distributions of $\gamma$}\label{subsec:GammaDistribution}

\begin{figure*}
    \begin{center}
        \includegraphics[width=0.8\textwidth]{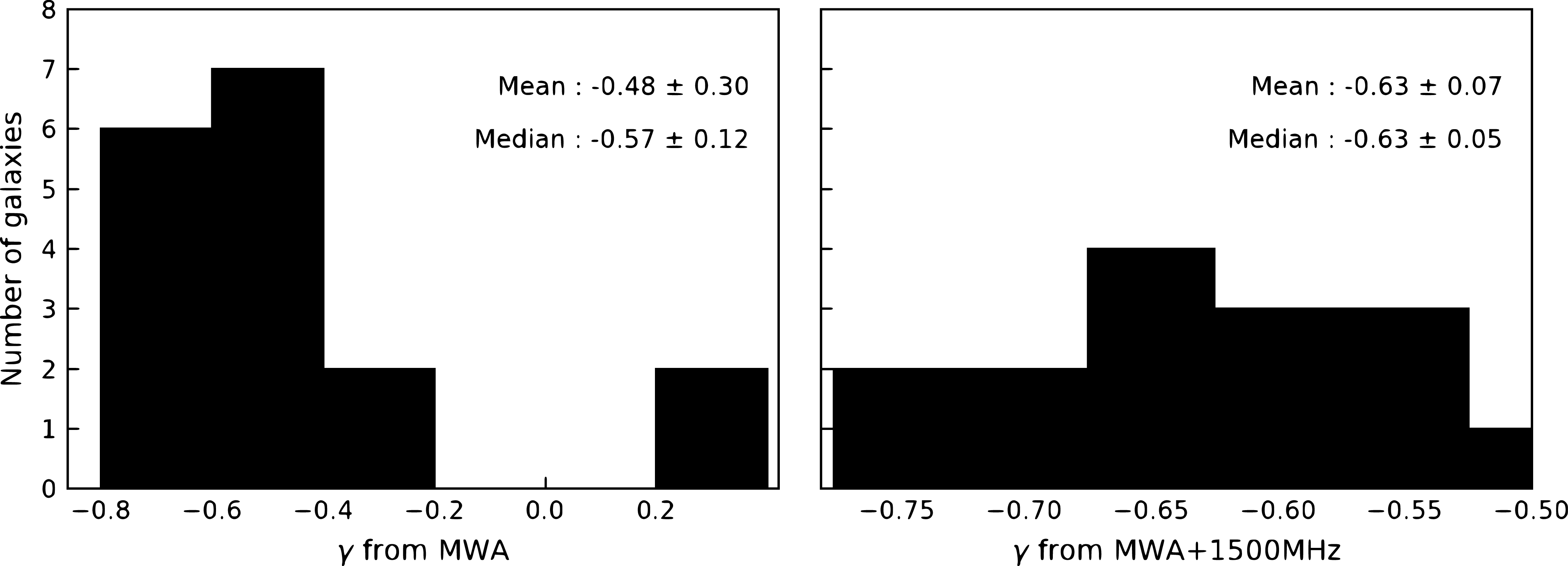}
    \end{center}
    \caption{
    Distribution of $\gamma$ in eq.~(\ref{eq:q_fitting}).
    Left: distribution of $\gamma$ obtained from fits to the MWA data only.
    Right: same as the left panel, but including the 1500~MHz flux density in the fit. 
    {Alt text: Histograms showing the distribution of the radio spectral index gamma derived from fits using MWA data alone (left) and MWA plus 1500 MHz data (right).} 
    }\label{fig:comparehist}
\end{figure*}

\begin{figure}
    \begin{center}
        \includegraphics[width=0.4\textwidth]{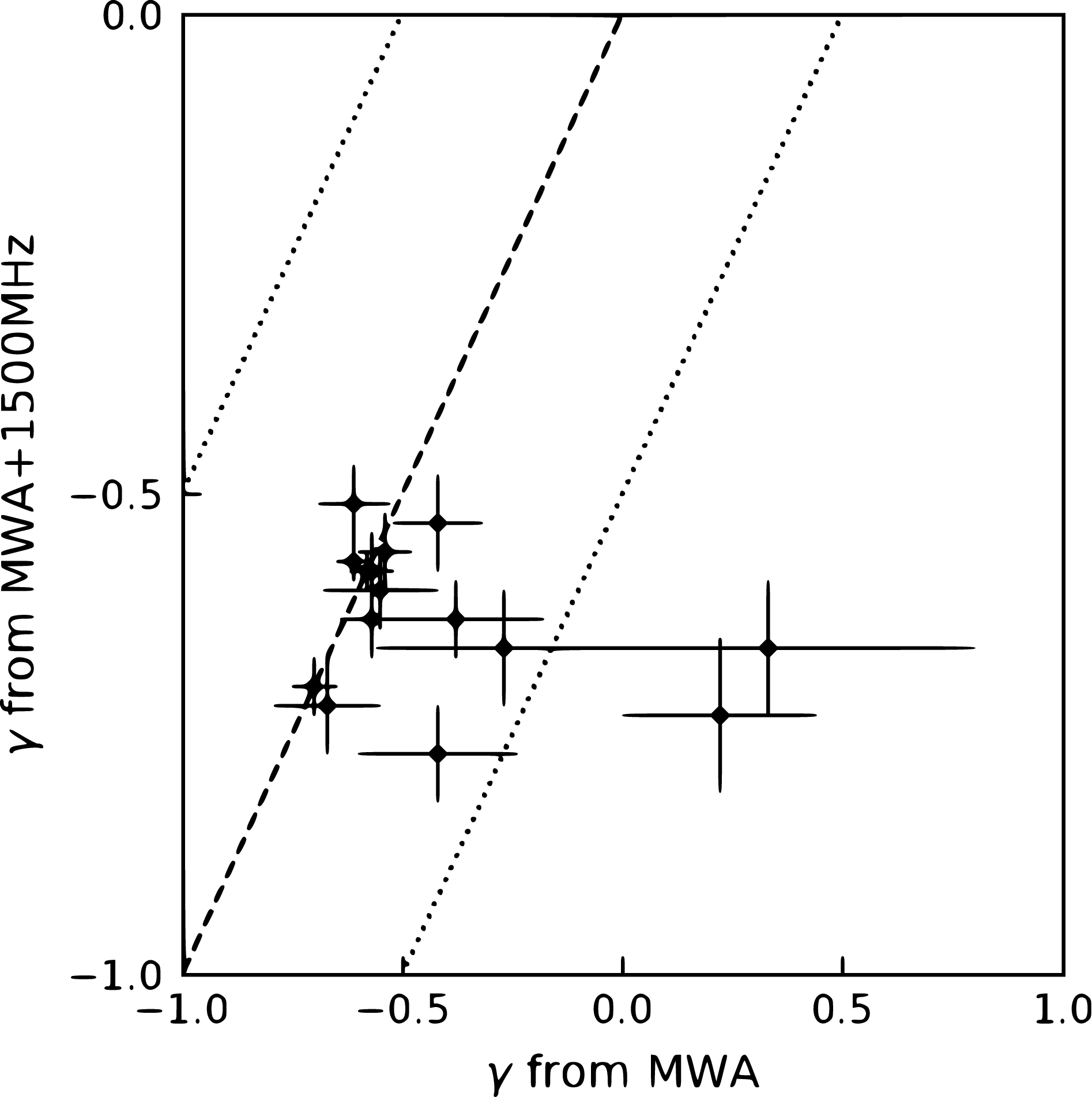}
    \end{center}
    \caption{
    Comparison between $\gamma$ obtained from MWA data with and without the 1500~MHz flux density.
    The error bars correspond to the uncertainties listed in Table~\ref{tab:estimated_properties}.
    The dashed line represents the 1-to-1 relation, and the dotted lines show an envelope obtained by shifting this line by 0.5 along both axes. 
    {Alt text: Scatter plot comparing the radio spectral index gamma obtained from MWA-only fits with those including the 1500 MHz data, with a one-to-one reference line.}
    }\label{fig:comparison_gamma}
\end{figure}

We first examine the distribution of the estimated spectral indices $\gamma$, shown in Figure~\ref{fig:comparehist}. 
The left panel presents the distribution derived from fits to MWA data alone, whereas the right panel shows the distribution obtained when the 1500~MHz data are included in the fit.
In both panels, we indicate the mean and median together with the standard deviation and interquartile range.

We note that three galaxies in our sample (HRS 122, 163, and 204) are fitted using only the MWA flux densities because their $1500\;\mbox{MHz}$ measurements have low signal-to-noise ratios \citep{Boselli2015}.

{
For the fits including the $1500\;\mbox{MHz}$ data, we obtain a median spectral index of $\gamma \simeq -0.63$ for the full sample.
Excluding the Seyfert/LINER galaxies HRS~144, HRS~163, and HRS~220 yields a very similar median value for the clean star-forming sample.
This result indicates that the ensemble spectral behavior is robust against the removal of the optically identified AGN hosts.
}

In the left panel of Figure~\ref{fig:comparehist}, two galaxies (HRS 25 and HRS 144) clearly stand out as outliers, whereas they do not show a significant deviation from the bulk of the sample once the 1500~MHz data are included.
The radio continuum measurements originate from the same compilation \citep{1990ApJS...73..359C} for all galaxies except HRS 306, whose flux is taken from \citet{2002AJ....124..675C}.
Therefore, systematic differences between catalogs are unlikely to explain these outliers, and indeed the estimated $\gamma$ for HRS 306 is consistent with the rest of the sample within the quoted uncertainty.

To investigate these outliers in more detail, we compare the values of $\gamma$ obtained with and without the 1500~MHz data on a galaxy-by-galaxy basis in Figure~\ref{fig:comparison_gamma}. 
The two outliers (HRS 25 and HRS 144) are clearly visible in this diagram.
We also see that, when the 1500~MHz flux density is included, the fitted slopes satisfy $\gamma < -0.5$ for all galaxies, even for those that showed flatter slopes when only MWA data were used.

This behavior suggests that a spectral turnover or curvature occurs at a frequency between the MWA band and 1500~MHz.
At low frequencies, the radio spectrum is expected to flatten because of free--free absorption \citep[e.g.,][]{CalistroRivera2017a, Schober2017, Chyzy2018}.
\citet{Schober2017} showed that a Milky Way-like galaxy (with a similar SFR) has a critical frequency roughly one order of magnitude lower than the MWA band.
However, HRS 25 and HRS 144 have SFRs comparable to that of the Milky Way \citep{Boselli2015}, yet appear to have a critical frequency between the MWA band and 1500~MHz, which is difficult to reconcile with the simple expectation of \citet{Schober2017}.

Several explanations are possible:
(1) limited statistics due to the small number of available flux-density measurements; and
(2) intrinsic physical peculiarity of the sources.

In the case of HRS 25, the flatter slope may largely reflect statistical uncertainty.
In contrast, HRS 144 is classified as a Seyfert galaxy based on the BPT diagram \citep[e.g.,][]{Baldwin1981, Kewley2001, Kauffmann2003, Schawinski2007}, implying that AGN-related radio emission may significantly affect the fitted spectral slope.
Furthermore, an H{\sc i} absorption feature toward the core of HRS 144 has been reported from the VIVA observations \citep{2009AJ....138.1741C}.

{
The clean-sample analysis therefore suggests that the ensemble spectral index distribution itself is relatively stable against the exclusion of the optically identified AGN hosts, although individual systems can still exhibit significant deviations from the ensemble behavior.
}

Additional broadband radio data would be required to establish the detailed origin of the observed spectral behavior in these systems.
The individual fitting results are presented in Appendix~\ref{sec:fitting_result}.

{
\subsection{Frequency dependence and scatter of the IR--radio correlation}
\label{subsec:sfrfromlowradio}
}

{
We next examine the ensemble frequency dependence of the IR--radio correlation itself and its galaxy-to-galaxy scatter through the reconstructed $q_\nu$ relation.
}

\begin{figure}
    \begin{center}
        \includegraphics[width=\columnwidth]{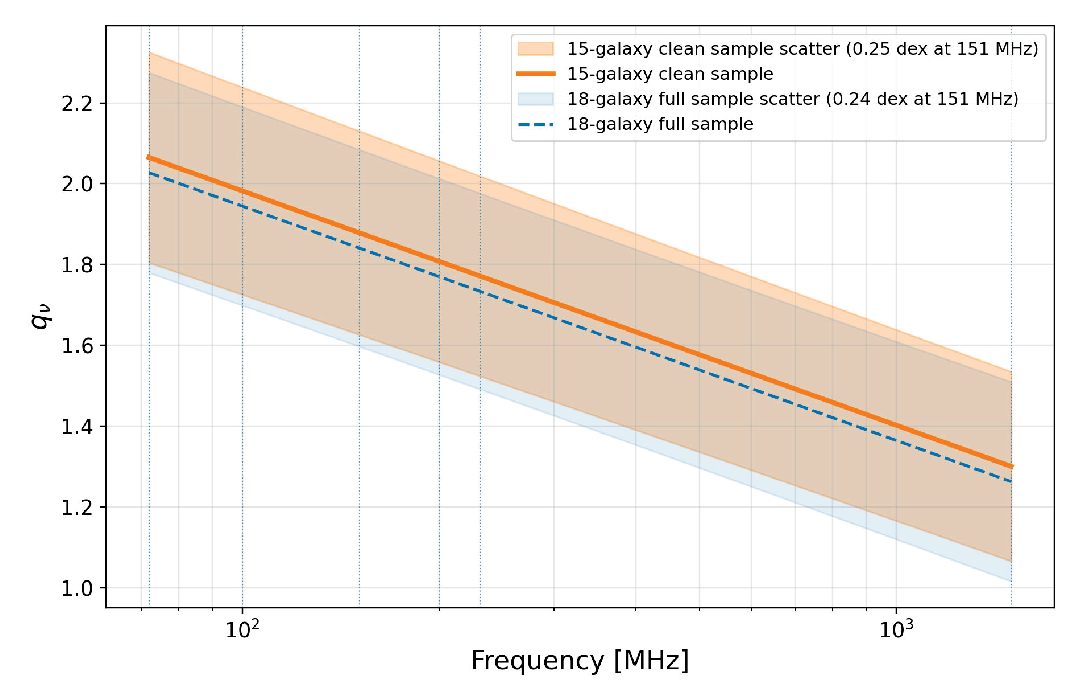}
    \end{center}
    \caption{
    {
    Frequency dependence of the IR--radio correlation coefficient $q_\nu$ of the sample.
    The solid orange line and shaded region show the median relation and scatter for the clean 15-galaxy sample, while the dashed blue line and shaded region show those for the full 18-galaxy sample.
    Vertical dotted lines indicate representative MWA frequencies and $1500\;\mbox{MHz}$.
    {Alt text: Frequency dependence of the IR--radio correlation coefficient q nu for the full and clean galaxy samples. The orange solid line and shaded band show the median relation and scatter for the clean 15-galaxy sample, while the blue dashed line and shaded band show those for the full 18-galaxy sample. The two relations are nearly identical, although substantial scatter remains in both samples across the MWA frequency range and up to 1500 MHz.}
    }
    }
    \label{fig:qnu_frequency_distribution}
\end{figure}

{
Figure~\ref{fig:qnu_frequency_distribution} shows the median $q_\nu$ relation as a function of frequency for both the full 18-galaxy sample and the clean 15-galaxy sample.
The median relations are nearly identical over the entire MWA frequency range and up to $1500\;\mbox{MHz}$.
This demonstrates that the ensemble frequency dependence of the IR--radio correlation itself is stable against the exclusion of the Seyfert/LINER galaxies.
}

{
At the same time, Figure~\ref{fig:qnu_frequency_distribution} shows that the galaxy-to-galaxy scatter remains substantial even after excluding the AGN-host systems.
To investigate this dispersion more directly, we examine the distribution of the measured $q_{151}$ values obtained from the spectral fitting results shown in Figure~\ref{fig:fitting_result}.
}

\begin{figure}
    \begin{center}
        \includegraphics[width=\columnwidth]{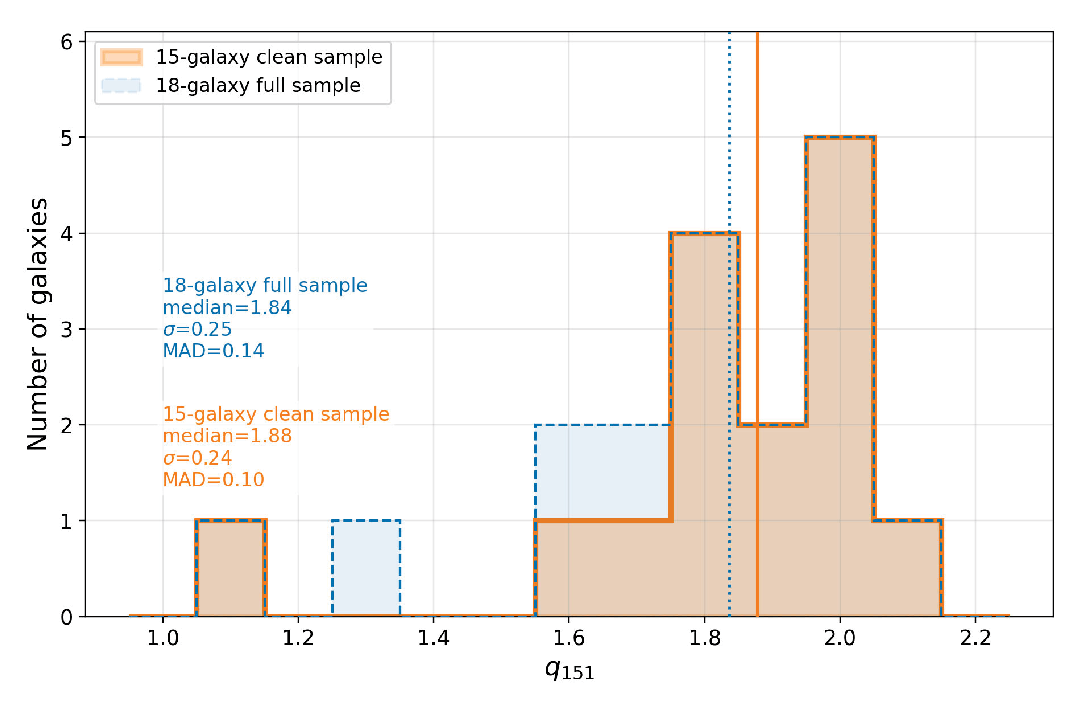}
        \includegraphics[width=\columnwidth]{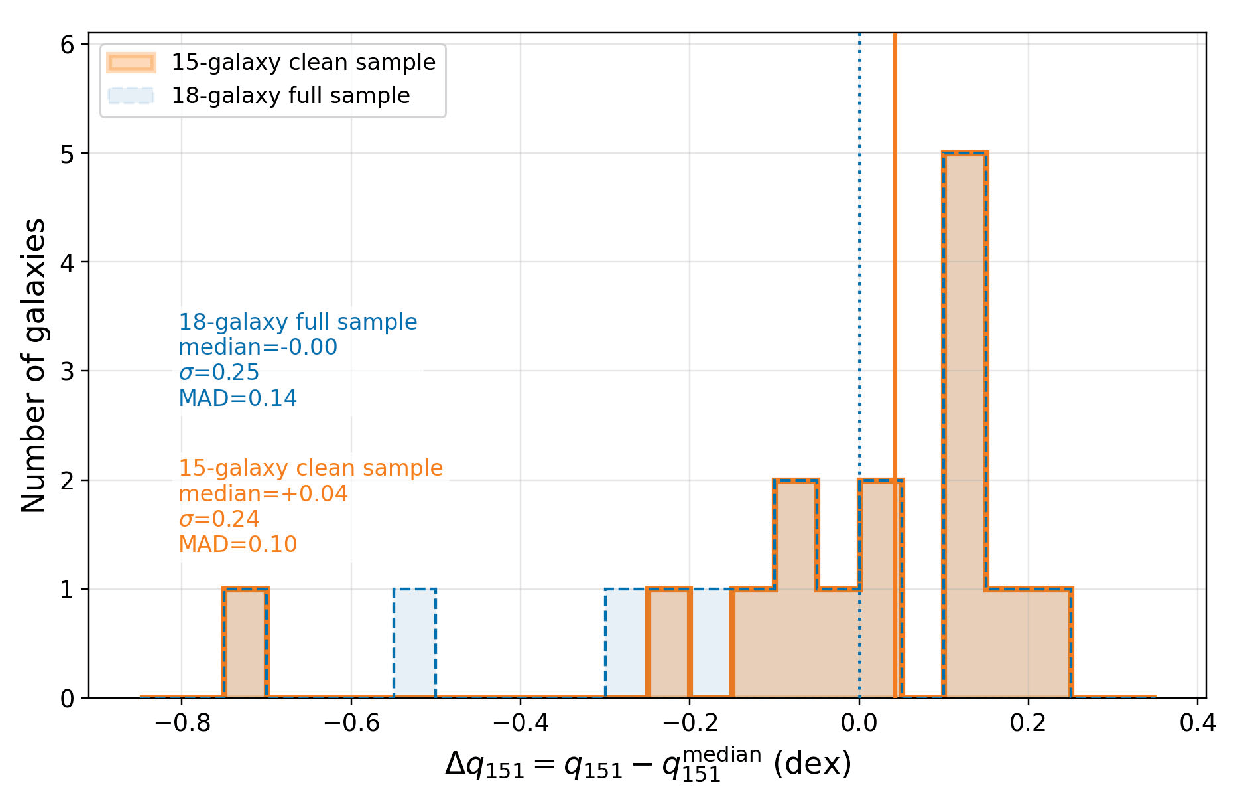}
    \end{center}
    \caption{
    {
    Top panel:
    Distribution of the measured $q_{151}$ values derived from the fitting results shown in Figure~\ref{fig:fitting_result}.
    The clean sample excludes the Seyfert/LINER galaxies HRS~144, HRS~163, and HRS~220.
    The clean sample exhibits a slightly narrower distribution than the full sample, although substantial intrinsic scatter remains in both cases.
    Bottom panel:
    Distribution of the offset parameter
    $\Delta q_{151} \equiv q_{151} - q_{151}^{\rm median}$
    for the full 18-galaxy sample (blue) and the clean 15-galaxy sample (orange).
    {Alt text:
    Two vertically stacked histograms showing the distribution of the low-frequency IR--radio correlation for the full 18-galaxy sample and the clean 15-galaxy sample.
    The top panel shows the measured q {151} values obtained from the spectral fitting results, while the bottom panel shows the corresponding offset parameter Delta q {151} relative to the ensemble median relation.
    In both panels, the clean sample excluding Seyfert/LINER galaxies exhibits only a modest reduction in scatter relative to the full sample, indicating that the residual dispersion of the low-frequency IR--radio relation cannot be explained solely by optically identified AGN contamination.}
    }
    }
    \label{fig:qnu_distribution}
\end{figure}

{
Figure~\ref{fig:qnu_distribution} shows that the median $q_{151}$ values for the clean and full samples are nearly identical, confirming that the ensemble low-frequency IR--radio relation itself is robust against the exclusion of the AGN-host galaxies.
At the same time, the scatter decreases only modestly:
$\sigma \simeq 0.25$ dex for the full sample and $\sigma \simeq 0.24$ dex for the clean sample.
This result indicates that the observed dispersion cannot be attributed solely to optically identified AGN contamination.
}

{
The residual scatter likely reflects a combination of several effects, including intrinsic radio excess associated with peculiar ISM conditions, environmental effects, variations in cosmic-ray transport, and observational uncertainties in the low-frequency radio luminosities.
In particular, HRS~306 remains a strong radio-excess outlier even within the clean sample, suggesting that non-AGN processes can also contribute substantially to the observed dispersion.
}

{
We emphasize that the radio-based SFRs derived in this work rely on IR-calibrated ensemble-averaged values of $q_\nu$ and $\gamma$.
Therefore, the primary result of this section is not the derivation of an independent SFR estimator for individual galaxies, but rather the characterization of the ensemble low-frequency IR--radio relation and its intrinsic scatter.
}

{
\subsection{Comparison with other SFR indicators}
\label{subsec:sfrcomparison}
}

\begin{figure}
    \begin{center}
    \includegraphics[width=0.4\textwidth]{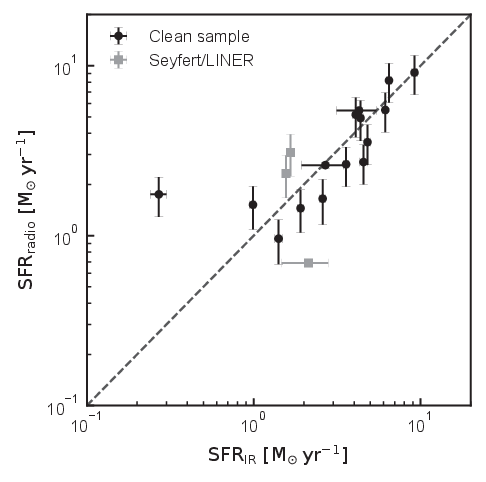}
    \includegraphics[width=0.4\textwidth]{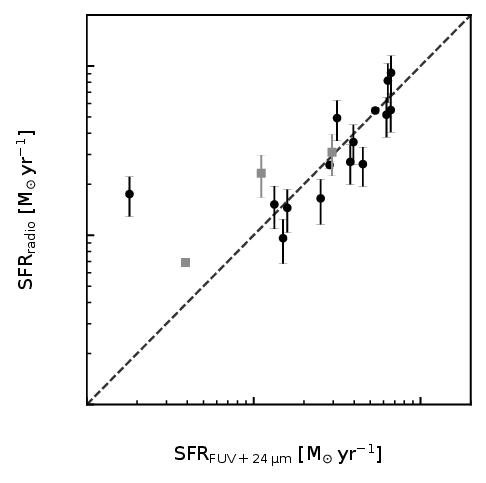}
    \includegraphics[width=0.4\textwidth]{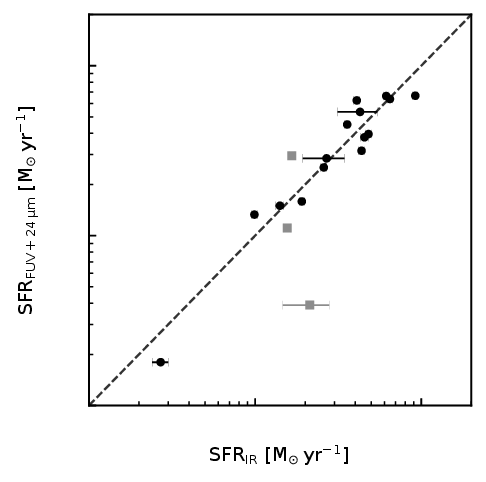}
    \end{center}
    \caption{
    Comparison between $\sfr_{\rm IR}$, $\sfr_{{\rm FUV+}24\mu{\rm m}}$, and $\sfr_{\rm radio}$. 
    From top to bottom, the panels show: 
    $\sfr_{\rm IR}$ vs.\ $\sfr_{\rm radio}$; 
    $\sfr_{{\rm FUV+}24\mu{\rm m}}$ vs.\ $\sfr_{\rm radio}$; and 
    $\sfr_{\rm IR}$ vs.\ $\sfr_{{\rm FUV+}24\mu{\rm m}}$.   
    {Alt text: Logarithmic comparisons between star formation rates derived from radio continuum, infrared emission, and UV plus mid-infrared hybrid indicators.}
    }
    \label{fig:SFR_ratio_comparison}
\end{figure}

{
Finally, we compare the radio-based SFRs with the IR- and UV-based SFR estimators.
}
Figure~\ref{fig:SFR_ratio_comparison} compares $\sfr_{\rm IR}$ and $\sfr_{\rm radio}$, $\sfr_{{\rm FUV+}24\mu{\rm m}}$ and $\sfr_{\rm radio}$, and $\sfr_{\rm IR}$ and $\sfr_{{\rm FUV+}24\mu{\rm m}}$. 
Here, $\sfr_{\rm radio}$ denotes the radio-based SFR at 151~MHz, $\mr{SFR}_{\mr{radio},\,151\,{\rm MHz}}$.

{
The top panel demonstrates that the radio SFR is, by construction, unbiased with respect to the IR-based calibration.
The remaining scatter reflects both the observational uncertainties in the radio luminosities and the intrinsic galaxy-to-galaxy diversity discussed in Section~\ref{subsec:sfrfromlowradio}.
}

Throughout this analysis, correlations and linear regressions are performed in logarithmic space.
The scatter of the correlation is comparable to the typical measurement uncertainties in the radio luminosities.
Thus, observational errors likely account for a substantial fraction of the dispersion.
At the same time, intrinsic variations among galaxies—such as differences in recent star-formation histories or ISM conditions—may also contribute at some level.

The middle panel of Figure~\ref{fig:SFR_ratio_comparison} presents the comparison between $\sfr_{{\rm FUV+}24\mu{\rm m}}$ and $\sfr_{\rm radio}$.
Again, the radio SFR estimator is broadly consistent with the UV+24\,$\mu$m hybrid estimator of \citet{2011ApJ...741..124H} used by \citet{Boselli2015}.
We note that \citet{Boselli2015} did not provide uncertainties for the UV+24\,$\mu$m estimates.

This behavior can be understood in terms of the characteristic timescales of the tracers.
UV-based SFR indicators averaged over continuous star formation probe timescales of order $10^7$--$10^8$ yr, depending on the assumed star-formation history.
\citet{2011ApJ...741..124H} showed observationally that the extinction-corrected H$\alpha$ and FUV luminosities are consistent with a constant SFR over $\sim100$ Myr \citep[see also][]{2003A&A...410...83H}.
The radio continuum emission traces star formation integrated over comparable or slightly longer timescales 
($\sim10^8$ yr), reflecting the lifetime and propagation of cosmic-ray electrons.
The similarity of these effective timescales likely contributes to the relatively tight correlation observed between 
$\sfr_{\rm radio}$ and $\sfr_{{\rm FUV+}24\mu{\rm m}}$ at moderate SFRs.

At higher SFRs, around $\sfr \simeq 10 \;\msun \,\mbox{yr}^{-1}$, the IR emission dominates the bolometric output from star formation \citep[e.g.,][]{2010A&A...514A...4T,2013EP&S...65..281T}.
In addition, as dust opacity increases at high SFR, the dust continuum emission shifts to longer wavelengths ($\lambda \gtrsim 60\;\mu{\rm m}$).
In this regime, the mid-IR emission is no longer a simple proxy for the total dust luminosity, and the 24\,$\mu$m-based hybrid SFR estimator tends to underestimate the true SFR \citep[e.g.,][]{2005A&A...432..423T}.
This behavior explains the larger deviations observed at high SFR in the middle panel of Figure~\ref{fig:SFR_ratio_comparison}.

The same effect is evident in the bottom panel of Figure~\ref{fig:SFR_ratio_comparison}, which compares $\sfr_{\rm IR}$ with $\sfr_{{\rm FUV+}24\mu{\rm m}}$.

{
Overall, the low-frequency radio continuum provides a robust ensemble tracer of star formation in nearby galaxies.
At the same time, the comparison with conventional IR- and UV-based tracers confirms that the intrinsic scatter of the low-frequency IR--radio relation remains non-negligible even after removing the optically identified AGN hosts.
}

\section{Discussion}\label{sec:discussion}

{
We now discuss the physical implications of the reconstructed low-frequency IR--radio relation and compare our results with previous low-frequency radio studies, particularly those based on LOFAR observations.
The principal result of this work is the stability of the ensemble frequency dependence of the low-frequency IR--radio correlation together with the persistence of non-negligible intrinsic scatter even after excluding the optically identified AGN hosts.
}

\subsection{Spectral index $\gamma$}

We first examine the value of the spectral index $\gamma$. 
Using the MWA and $1500\,\mathrm{MHz}$ data, we obtained a median spectral index of $\gamma \simeq -0.63$, while \citet{CalistroRivera2017a} and \citet{Chyzy2018} reported $-0.78 \pm 0.24$ and $-0.56 \pm 0.11$, respectively.
All of these values are mutually consistent within the quoted uncertainties.

The modest differences among studies are most likely driven by differences in sample selection.
Our sample consists of nearby galaxies within $25\;\mathrm{Mpc}$, whereas \citet{CalistroRivera2017a} analyzed 758 galaxies up to $z\sim 2$, and \citet{Chyzy2018} used 118 galaxies up to $z = 0.04$.
Because higher-redshift samples probe higher rest-frame frequencies, the observed spectral slopes may become steeper through a combination of intrinsic spectral curvature and redshift effects \citep[e.g.,][]{2011A&A...529A..22B,2011ApJ...731...79M,2013A&A...556A.142S,2016A&A...592A.155S,2021A&A...647A.123D,2021ApJ...914..126M}. 
Indeed, \citet{Chyzy2018} reported steeper slopes at frequencies of $1.3\mbox{--}5\;\mathrm{GHz}$, suggesting that broadband measurements can capture spectral curvature associated with synchrotron aging or free--free absorption.

Recent broadband radio studies have shown that star-forming galaxies often exhibit low-frequency spectral flattening or turnovers below $\sim200$ MHz \citep[e.g.,][]{2024MNRAS.528.5346A,2025PASA...42....2G}.
Our result, based on 20 sub-bands between 72 and 231~MHz, shows no strong evidence for systematic curvature within this frequency range for the majority of our galaxies, supporting the use of a single power-law approximation for nearby star-forming systems.

{
At the same time, the clean-sample analysis presented in Section~\ref{subsec:GammaDistribution} demonstrates that the ensemble spectral-index distribution itself is remarkably stable against the exclusion of the optically identified AGN hosts.
This suggests that the median spectral behavior is largely controlled by the star-forming galaxy population rather than by a small number of AGN-contaminated systems.
}

However, individual galaxies can still deviate significantly from the ensemble relation.
In particular, HRS~25 and HRS~144 exhibit flatter spectral behavior when only the MWA data are used.
For HRS~144, the presence of Seyfert activity and H{\sc i} absorption strongly suggests that nuclear activity influences the radio spectrum.
Larger samples with deeper broadband radio data will be required to determine whether similar spectral behavior is common among specific galaxy populations or environments.

{
\subsection{Scatter and stability of the low-frequency IR--radio relation}
}

{
The reconstructed $q_\nu$ relation presented in Section~\ref{subsec:sfrfromlowradio} reveals two important features.
First, the median frequency dependence of the low-frequency IR--radio relation is highly stable against the exclusion of the Seyfert/LINER galaxies.
Second, the galaxy-to-galaxy scatter remains substantial even for the clean sample.
}

{
The clean sample yields a scatter of $\sigma \simeq 0.25$ dex, only slightly smaller than the $\sigma \simeq 0.27$ dex obtained for the full sample.
This result indicates that the residual dispersion cannot be explained solely by classical AGN contamination identified from optical diagnostics.
Instead, the scatter likely reflects a combination of several physical effects, including environmental dependence of synchrotron emission, variations in cosmic-ray transport, magnetic-field amplification, and intrinsic radio excess associated with peculiar ISM conditions.
}

{
In particular, HRS~306 remains a strong radio-excess outlier even after excluding the Seyfert/LINER systems.
This suggests that significant deviations from the ensemble low-frequency IR--radio relation can arise even in systems without obvious optical AGN signatures.
The persistence of such outliers implies that the intrinsic scatter of the low-frequency IR--radio relation is itself an important physical observable, rather than merely a by-product of AGN contamination.
}

{
Our results therefore support the view that the low-frequency IR--radio relation is best interpreted statistically.
The ensemble relation itself appears stable and robust, while the galaxy-to-galaxy scatter encodes additional information about ISM conditions, cosmic-ray propagation, and environmental effects.
}

\subsection{Implications for radio-based SFR calibrations}

We next compare our global radio-SFR calibration with those presented in the literature.
Substituting the median values $\gamma = -0.63$, $q_{1500\,[{\rm MHz}]} = 2.48$, and $\nu = 150\,\mathrm{MHz}$ into eq.~(\ref{eq:sfrfromradio}) yields
\begin{align}
  \sfr_{\rm radio,151\,MHz}
  =
  \left(1.03\pm0.27\right)\times10^{-29}
  L_{\rm radio,151\,MHz}
  \;[\mathrm{erg\,s^{-1}}] .
\end{align}

For comparison, \citet{CalistroRivera2017a} reported a coefficient of $0.76 \pm 0.08$ at $z=0$.
This difference is well within the expected uncertainties associated with sample selection and intrinsic scatter.

Several LOFAR-based studies have examined the relation between low-frequency radio luminosity and SFR.
\citet{2018MNRAS.475.3010G} studied 150~MHz luminosities for galaxies at $0 \lesssim z \lesssim 0.3$ in the {\sl Herschel} H-ATLAS NGP field and found a nearly linear radio--SFR relation.
Similarly, \citet{2019A&A...631A.109W} and \citet{2021A&A...648A...6S} reported relations broadly consistent with linear scaling between low-frequency radio luminosity and SFR over a wide range of redshift and galaxy properties.

{
Our analysis differs from these studies in that the primary focus is not the calibration coefficient itself, but rather the stability and scatter of the ensemble low-frequency IR--radio relation.
The reconstructed $q_\nu$ relation demonstrates that low-frequency radio luminosity provides a robust ensemble tracer of star formation activity, while simultaneously showing that non-negligible intrinsic scatter remains even after excluding optically identified AGN hosts.
}

Although most published relations express radio luminosity as a function of SFR rather than the reverse, the nearly linear nature of the relation makes comparison straightforward.
Overall, our radio-SFR calibration is fully consistent with the bulk of LOFAR-based studies at $z \simeq 0$.

{
We also note that the HRS sample is approximately stellar-mass selected through the $K$-band criterion.
Recent studies suggest that incorporating stellar mass may reduce the scatter in radio-based SFR indicators.
A detailed investigation of such secondary dependencies is beyond the scope of the present work, but will be an important subject for future studies with larger low-frequency samples.
}

\section{Summary}\label{sec:summary}

In this study, we investigated the low-frequency infrared (IR)--radio correlation for a sample of 18 nearby galaxies drawn from the {\sl Herschel} Reference Survey (HRS), using radio continuum flux densities obtained with the Murchison Widefield Array (MWA) as part of the GLEAM survey.

We focused on the frequency dependence of the IR--radio correlation coefficient $q_\nu$, modeling its variation as
\begin{align}
  q_\nu = -\gamma \log \nu + \beta \nonumber
\end{align}
[eq.~(\ref{eq:q_fitting})].
Using the densely sampled GLEAM sub-bands between 72 and 231~MHz together with the $1500$~MHz measurements, we found that the ensemble low-frequency IR--radio relation is well described by a single power-law form with a median spectral index of $\gamma \simeq -0.63$ and $q_{1500\,{\rm MHz}} \simeq 2.48$.

{
The reconstructed $q_\nu$ relation remains remarkably stable even after excluding the Seyfert/LINER galaxies from the sample, indicating that the ensemble spectral behavior is not strongly driven by the optically identified AGN hosts.
}

{
At the same time, the galaxy-to-galaxy scatter of the low-frequency IR--radio relation remains substantial.
To quantify this dispersion, we examined the distribution of the measured $q_{151}$ values obtained from the spectral fitting (Fig.~4) and defined the offset parameter
}
\begin{align}
  {
  \Delta q_{151} \equiv q_{151} - q_{151}^{\rm median}.
  }
\end{align}

{
We found that the clean 15-galaxy sample exhibits a scatter of $\sigma \simeq 0.24$ dex, only slightly smaller than the $\sigma \simeq 0.25$ dex obtained for the full 18-galaxy sample.
This result suggests that the observed dispersion cannot be explained solely by classical AGN contamination, and that additional physical effects (such as environmental dependence, variations in cosmic-ray transport, and intrinsic radio excess) likely contribute to the scatter of the low-frequency IR--radio relation.
}

We also examined the consistency of the radio-based star formation rate estimator, $\sfr_{\rm radio}$, with IR- and UV-based SFR tracers.
The radio SFR estimator is broadly consistent with conventional tracers and behaves stably over the MWA frequency range, although a non-negligible intrinsic scatter remains.

{
These results support the use of low-frequency radio continuum as a robust ensemble tracer of star formation activity in nearby galaxies, while highlighting the need to explicitly account for the intrinsic scatter of the low-frequency IR--radio relation.
}

A more comprehensive understanding of low-frequency radio emission and its connection to star formation will require larger samples with broadband radio data and improved angular resolution.
In this regard, the forthcoming upgraded GLEAM survey {(GLEAM-X: \citealt{2022PASA...39...35H, 2024PASA...41...54R})} will be particularly promising, offering significant improvements in both sensitivity and spatial resolution relative to the current dataset.
Additional constraints will also come from new wide-area radio surveys such as RACS \citep{2021PASA...38...58H}.
We leave exploration of these expanded datasets for future work.

\begin{ack}

{
First of all, we thank the referees for their valuable constructive comments that greatly improved the clarity of this paper.}

This work is based on the HRS data together with GLEAM survey maps, both of which are available from their data archive in public.  

This work has been supported by the Japan Society for the Promotion of Science (JSPS) Grants-in-Aid for Scientific Research (19H05076, 21H01128, and 24H00247). 
This work has also been supported in part by the Sumitomo Foundation Fiscal 2018 Grant for Basic Science Research Projects (180923), and the Collaboration Funding of the Institute of Statistical Mathematics ``New Development of the Studies on Galaxy Evolution with a Method of Data Science''. 
SC has been supported by the JSPS (21J23611). 
Parts of this research were conducted by the Australian Research Council Centre of Excellence for All Sky Astrophysics in 3 Dimensions (ASTRO 3D), through project number CE170100013.

\end{ack}

\bibliographystyle{apj}
\bibliography{radio_continuum_SFR}
%\bibliography{Updated_Radio_continuum_SFR}

\appendix

\section{Galaxy images}\label{sec:galaxy_image}

Here we show the radio contours superposed on the SDSS $i$-band images of all the sample galaxies in Figure~\ref{fig:galaxy_image}.

\begin{figure*}
    \begin{center}
        \includegraphics[width=0.9\textwidth]{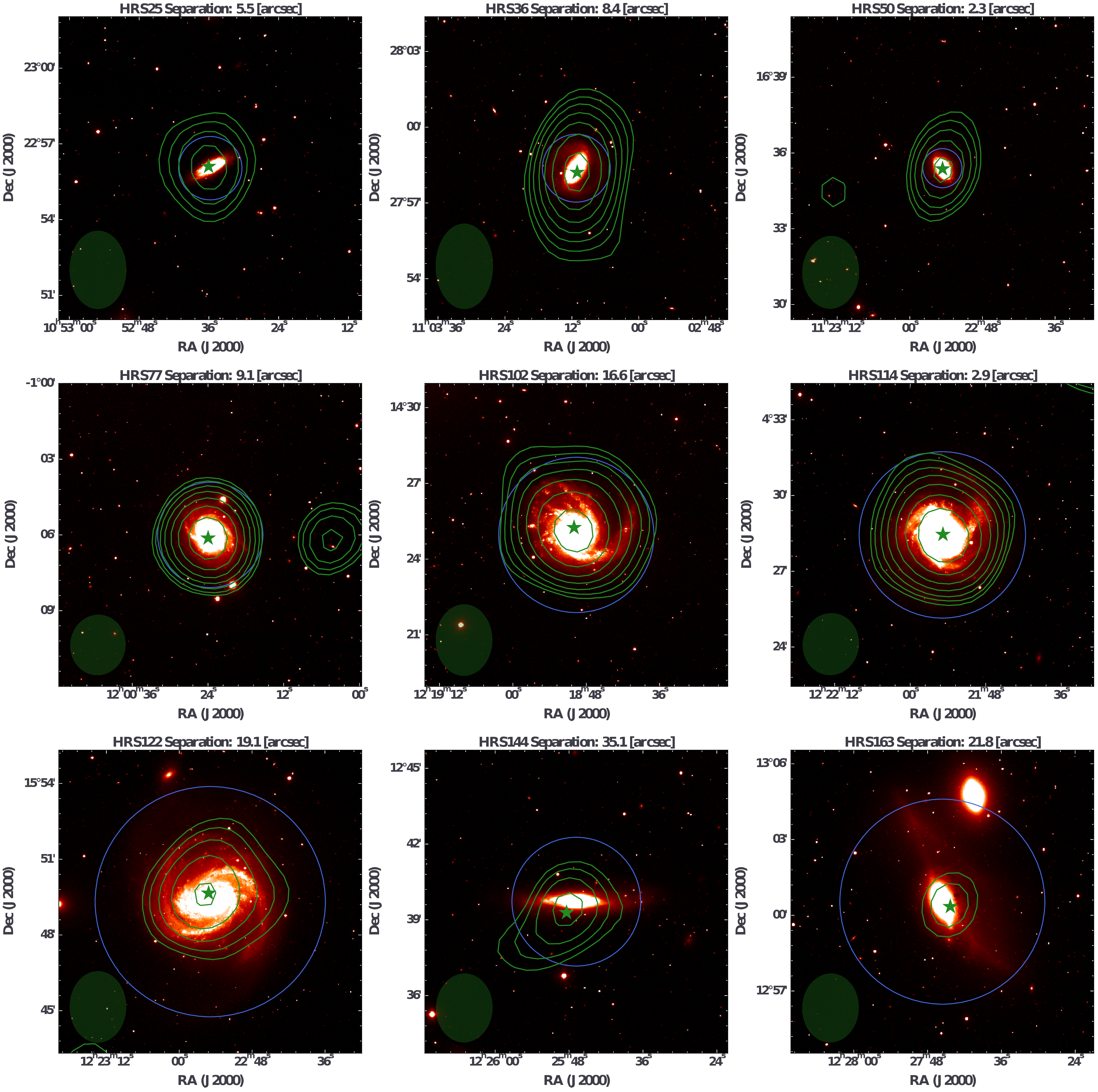}
    \end{center}
    \caption{{
    These are the SDSS \citep{Abolfathi2018} $i$-band images with low-frequency radio contours from the GLEAM survey at $170\mbox{--}231$~MHz.
    The radio contours are shown as solid lines, starting from the $3\sigma$ local noise level and increasing by a factor of $\sqrt{2^n}$, where $n$ is an integer.
    The star marker indicates the position of the corresponding source in the GLEAM catalog; the marker size has no physical or observational meaning.
    The circle indicates the isophotal optical size at $25\,\mr{mag}\arcs^{-2}$ from \citet{2010PASP..122..261B}.
    The ellipse in the lower-left corner of each panel shows the GLEAM beam size.
    These galaxies are selected in Section~\ref{sec:sample}.
    {Alt text: Optical SDSS i-band images of the sample galaxies with MWA radio continuum contours overlaid, illustrating the spatial correspondence between stellar emission and low-frequency radio emission.}}
    }\label{fig:galaxy_image}
\end{figure*}

\setcounter{figure}{5}
\begin{figure*}
    \begin{center}
    \includegraphics[width=0.9\textwidth]{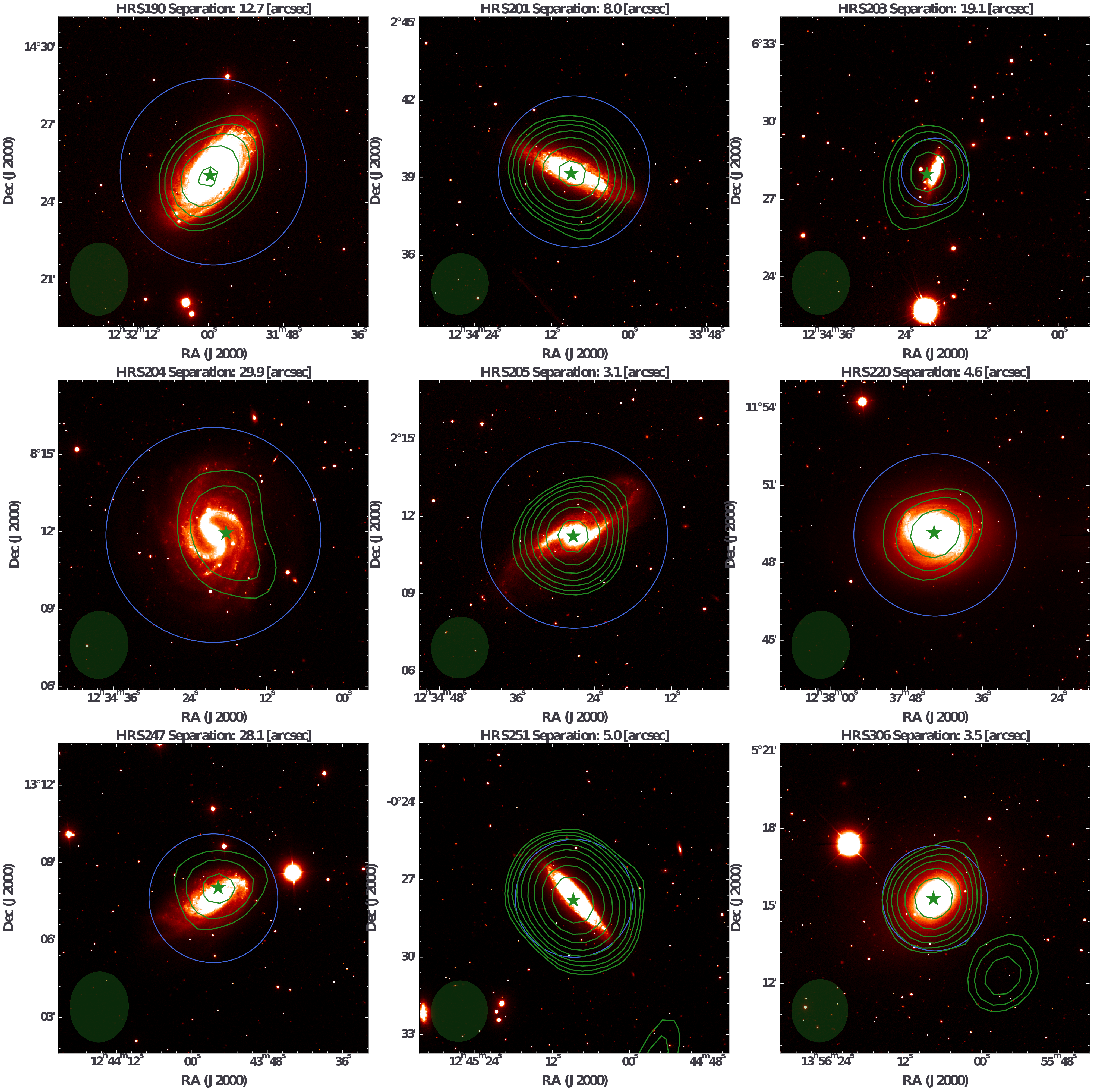}
    \end{center}
    \caption{
    --- Continued.
     {Alt text: Optical SDSS i-band images of the sample galaxies with MWA radio continuum contours overlaid, illustrating the spatial correspondence between stellar emission and low-frequency radio emission.}
    }\label{fig:galaxy_image2}
\end{figure*}

\newpage

\section{Fitting results}\label{sec:fitting_result}

\begin{figure*}
    \begin{center}
    \includegraphics[width=0.9\textwidth]{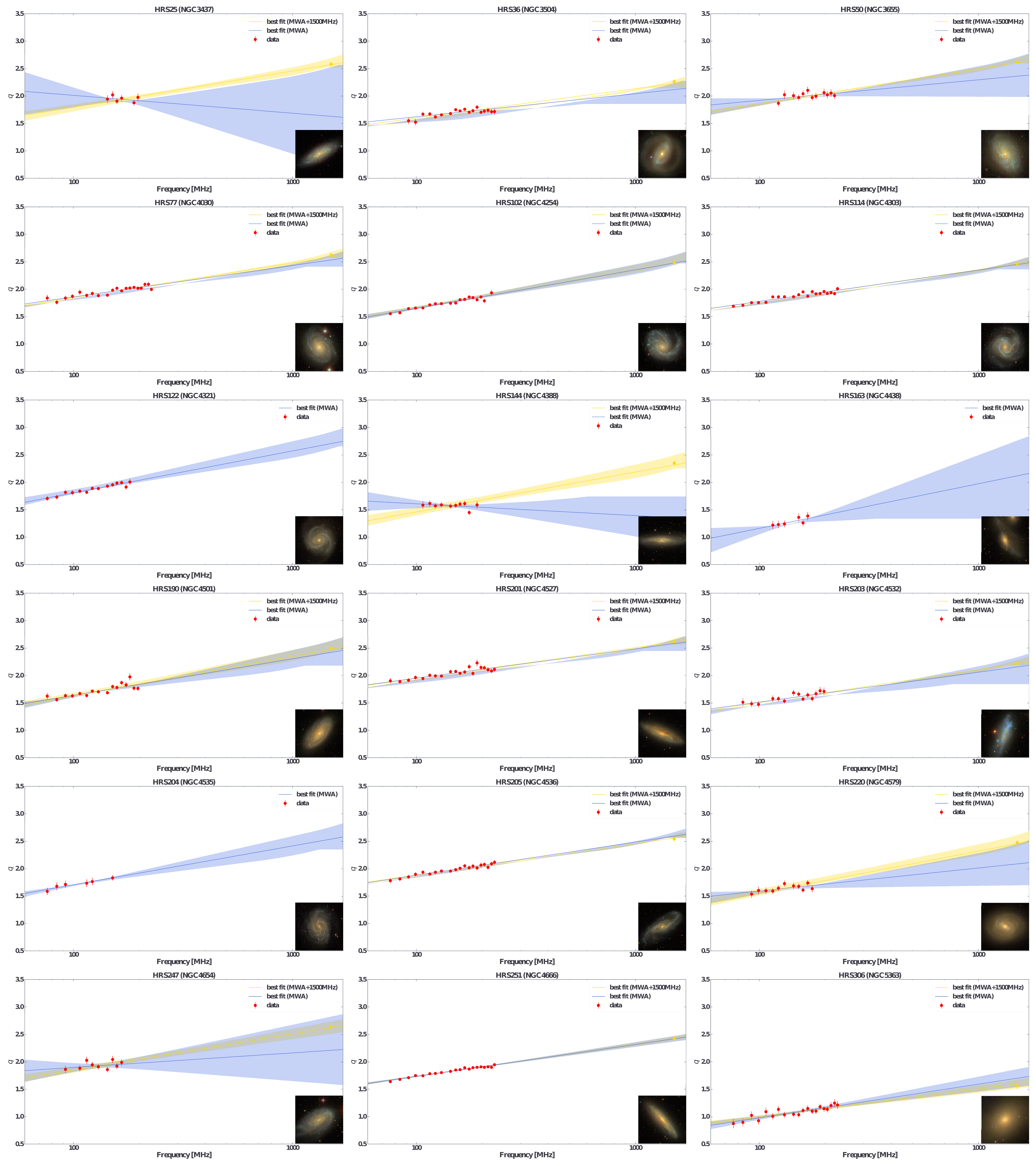}
    \end{center}
    \caption{{
    The fitting results.
    Filled circles indicate the flux densities measured in the MWA sub-bands, and the single point at 1500 MHz is taken from \citet{Boselli2015}.
    The solid line represents the best-fitting relation obtained using only the MWA data, and the surrounding shaded region shows its 95~\% confidence interval.
    The second fitted relation, obtained by including the 1500 MHz measurement, is also shown together with its corresponding 95~\% confidence interval.
    For HRS122, HRS163, and HRS204, this second fit is omitted because no high-quality 1500 MHz measurement is available.
    The SDSS RGB composite image of each galaxy is shown in the lower-right corner of each panel.
    {Alt text: Radio spectral energy distribution of an individual HRS galaxy, showing measured flux densities across the MWA bands together with the best-fitting power-law models based on the MWA data alone and on the combined MWA+1500 MHz data.}
    }}\label{fig:fitting_result}
\end{figure*}

The fitting results of the whole sample is presented in Figure~\ref{fig:fitting_result}.

\end{document}